\documentclass{iaus}
\usepackage{graphicx}

  \checkfont{eurm10}
  \iffontfound
    \IfFileExists{upmath.sty}
      {\typeout{^^JFound AMS Euler Roman fonts on the system,
                   using the 'upmath' package.^^J}%
       \usepackage{upmath}}
      {\typeout{^^JFound AMS Euler Roman fonts on the system, but you
                   dont seem to have the}%
       \typeout{'upmath' package installed. iaus.cls can take advantage
                 of these fonts,^^Jif you use 'upmath' package.^^J}%
      }
  \else
  \fi

  \checkfont{msam10}
  \iffontfound
    \IfFileExists{amssymb.sty}
      {\typeout{^^JFound AMS Symbol fonts on the system, using the
                'amssymb' package.^^J}%
       \usepackage{amssymb}%

      }{}
  \fi

  \IfFileExists{amsbsy.sty}
    {\typeout{^^JFound the 'amsbsy' package on the system, using it.^^J}%
     \usepackage{amsbsy}}
    {}

\newsavebox{\astrutbox}
\sbox{\astrutbox}{\rule[-5pt]{0pt}{20pt}}

\newcommand{\um}{$\,\mu$m}

\title[The Interplay among Black Holes, Stars and ISM in Galactic 
       Nuclei]{Infrared Observations of AGN}

\author[P. Lira {\it et al.\/}]%
{P. Lira$^1$, F. Mar\'{\i}n$^{1}$, L. Videla$^1$,\break
A.~Alonso-Herrero$^2$, D.~M.~Alexander$^3$ \and M.~Ward$^4$}

\affiliation{
$^1$ Universidad de Chile, Casilla 36-D, Santiago, Chile\\[\affilskip]
$^2$ DAMIR - IEM, C.~Serrano 113-121, 28006 Madrid, Spain\\[\affilskip]
$^3$ Cambridge University, Madingley Road, Cambridge CB3~0HA, UK\\[\affilskip]
$^4$ Leicester University, Leicester LE1~7RH, UK}

\pubyear{2004}
\volume{222}
\pagerange{1--8}
\date{?? and in revised form ??}
\jname{The Interplay among Black Holes, Stars and ISM \\in Galactic Nuclei}
\editors{Th. Storchi Bergmann, L.C. Ho \& H.R. Schmitt, eds.}
\begin{document}

\maketitle

\begin{abstract}
We present results from an imaging and spectroscopic study of the dust
properties of Seyfert galaxies in the 1-10\um\ range. The data are
compared to state of the art models of torus emission to constrain
geometrical and physical properties of the obscuring medium.
\end{abstract}

\firstsection 

\section{Introduction}

A key ingredient of the Unified Model for AGN is an axially symmetric
dusty structure (a.k.a., the torus) that absorbs UV photons from the
central source and re-emits them anisotropically as thermal radiation
at near and mid-IR wavelengths. This wavelength range is, therefore,
crucial to determine the properties of the obscuring material.

Here we present 1-10\um\ nuclear SEDs and N-band spectroscopy for a
small sample of Seyfert galaxies. This is part of a much larger study
of Seyfert galaxies drawn from the 12\um\ Extended Galaxy Sample
(Rush, Malkan \& Spinoglio, 1993) for which we are collecting
JHKLMN-band imaging in order to produce nuclear SEDs. The results will
be compared with theoretical models of the dusty torus (Nenkova,
Ivezi\' c \& Elitzur, 2002), observations of the N$_{\rm H}$ column
obtained from X-ray observations (Bassani et~al., 1999), and
spectropolarimetry (Tran, 2003). We expect to put statistically
significant constraints on the physical and geometrical
characteristics of the obscuring material around AGN.

\section{Observations, data analysis and future work}

\begin{figure}
\centering
\includegraphics[angle=270,scale=0.28]{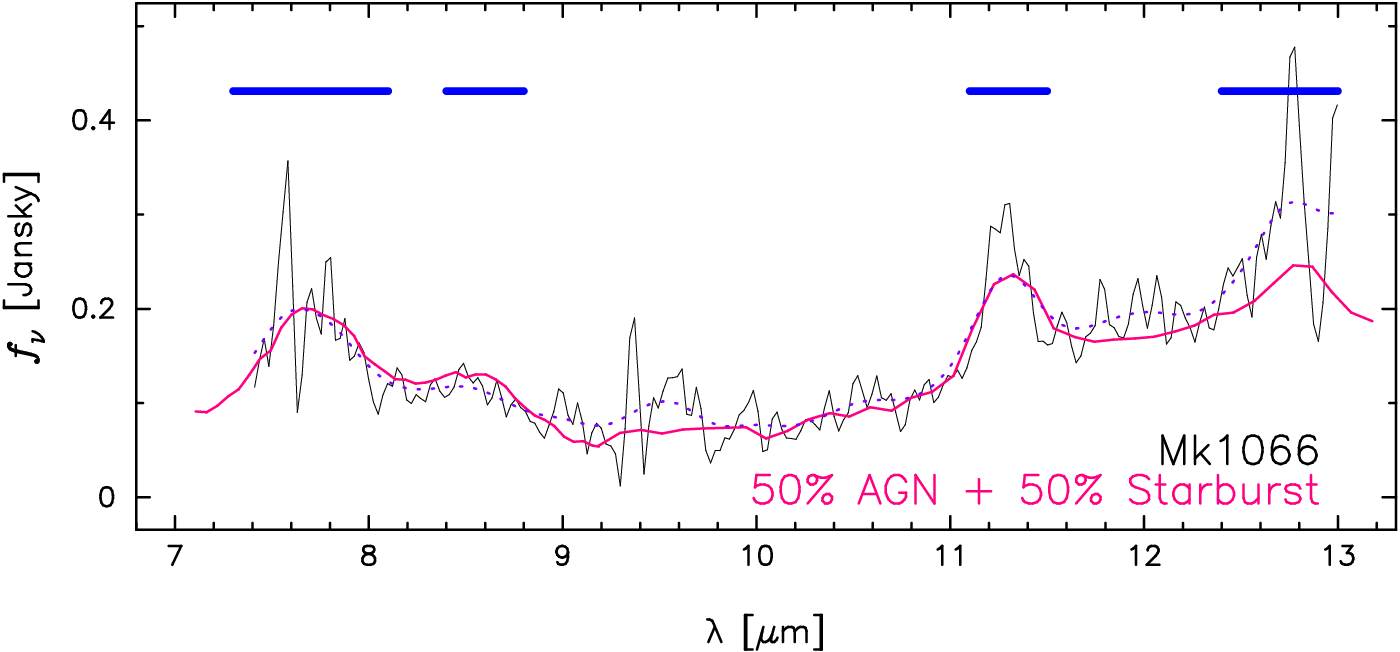}\hfil
\includegraphics[angle=270,scale=0.30]{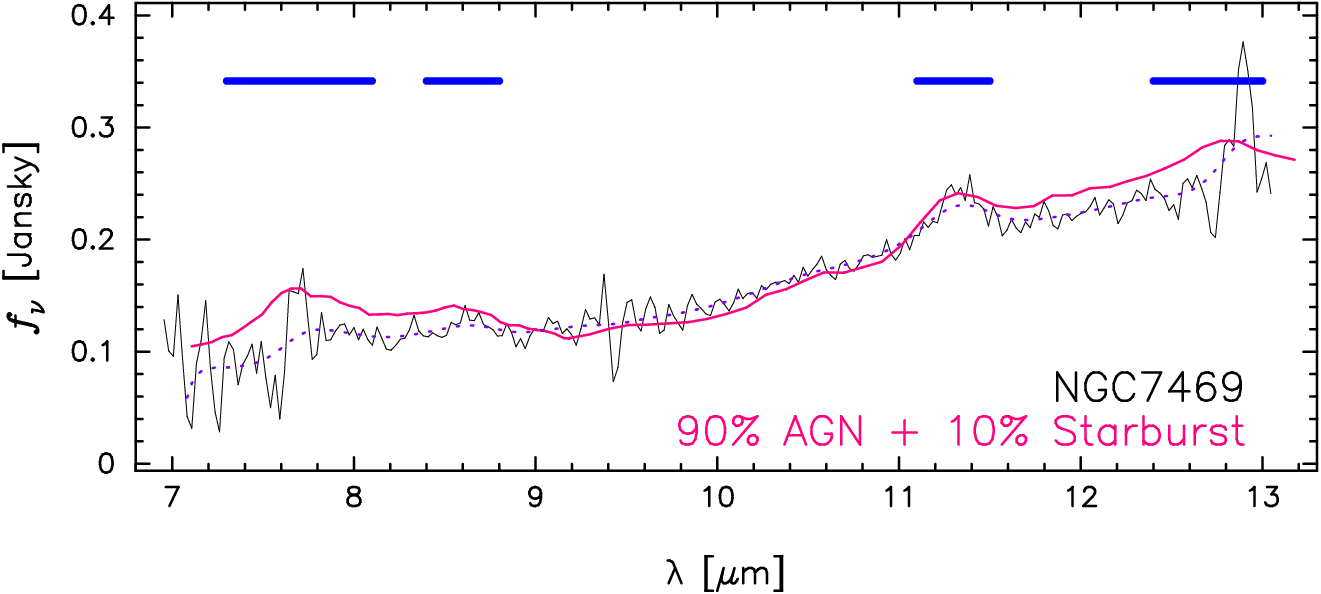}
\caption{N-band Michelle spectra for 2 of the Seyfert galaxies in our
sample (thin continuous line). The positions of the more prominent
PAHs are shown. The spectra were smoothed (dashed line) for comparison
with ISO templates of AGN $+$ starburst emission in order to estimate
the contribution to the data from star-forming regions (thick
continuous line).}
\end{figure}

We have obtained UKIRT/Michelle N-band spectroscopy for 4 Seyfert 2s
(S2s: Mk509, Mk1066, NGC1068, NGC7674) and 2 Seyfert 1 galaxies (S1s:
MCG+8-11-11, NGC7469). Using a PAH dominated template from starburst
emission (kindly provided by E.~Le Floc'h) we have estimated the
contribution to the observed spectra from circumnuclear star-forming
regions. We find that only for Mk1066 the starburst contribution is
significant ($\sim 50\%$), being typically $\lesssim 10\%$ for all
other nuclei (Fig.~1). Isolating the AGN continuum is crucial when
trying to determine the profile of the 9.7\um\ dust feature, which can
be very useful as a diagnistic for dust conditions (Nenkova, Ivezi\' c
\& Elitzur, 2002), since the starburst related PAHs can mimic a dip at
smimilar wavelengths. We find that the 9.7\um\ feature is probably
present in all spectra.

We then compared the N-band spectra and nuclear 1-10\um\ SEDs
determined for our targets (Alonso-Herrero et~al., 2003), to derive
constraints on the obscuring material around the central engine. In
the examples shown in Fig.~2 we find that the models provide a good
description of the IR data for Mk573 and a satisfactory fit for
NGC7674, where the photometric data is well matched by the model, but
the 9.7\um\ dust feature is not well reproduced. The high inclination
angle of the torus in NGC7674 is in good agreement with the
Compton-Thick column obseved in X-rays (Bassani et~al., 1999).

\begin{figure}
\centering
\includegraphics[angle=270,scale=0.29]{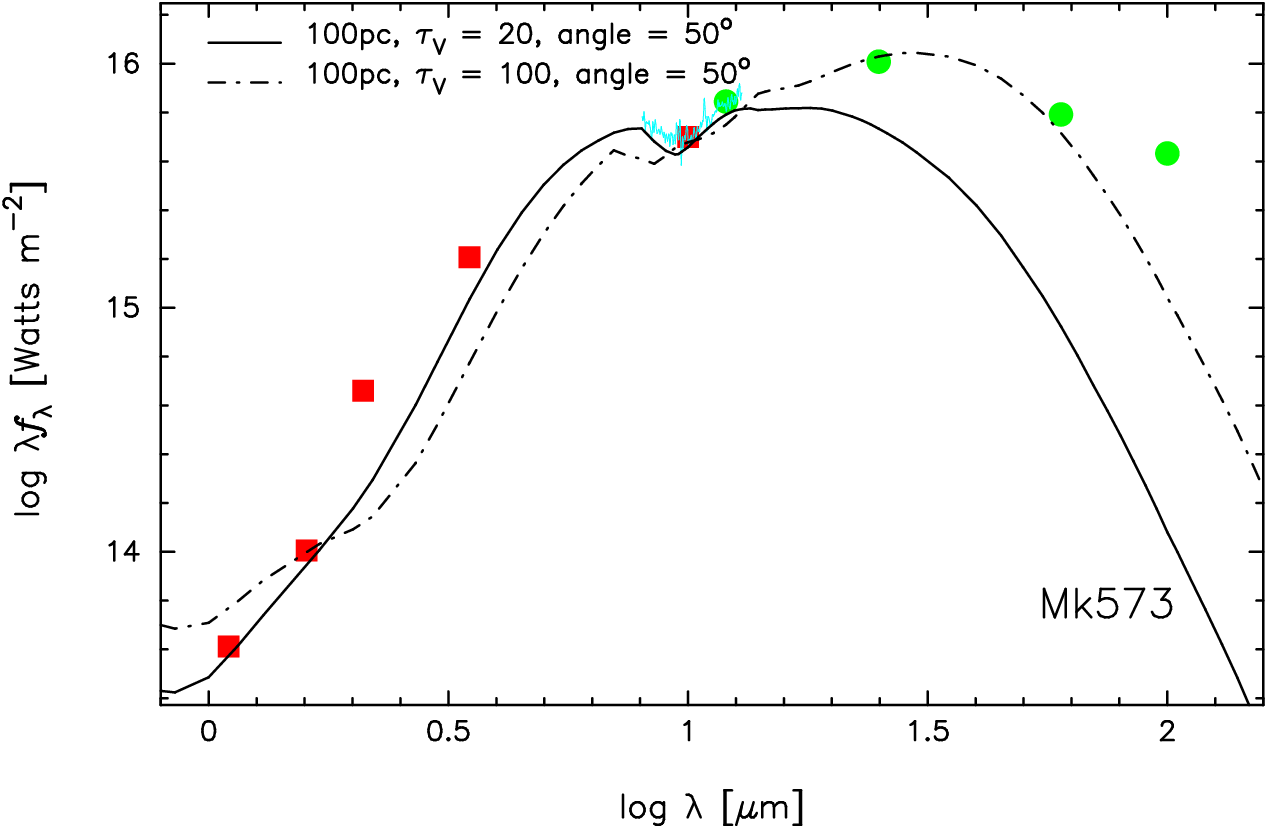}\hfil
\includegraphics[angle=270,scale=0.29]{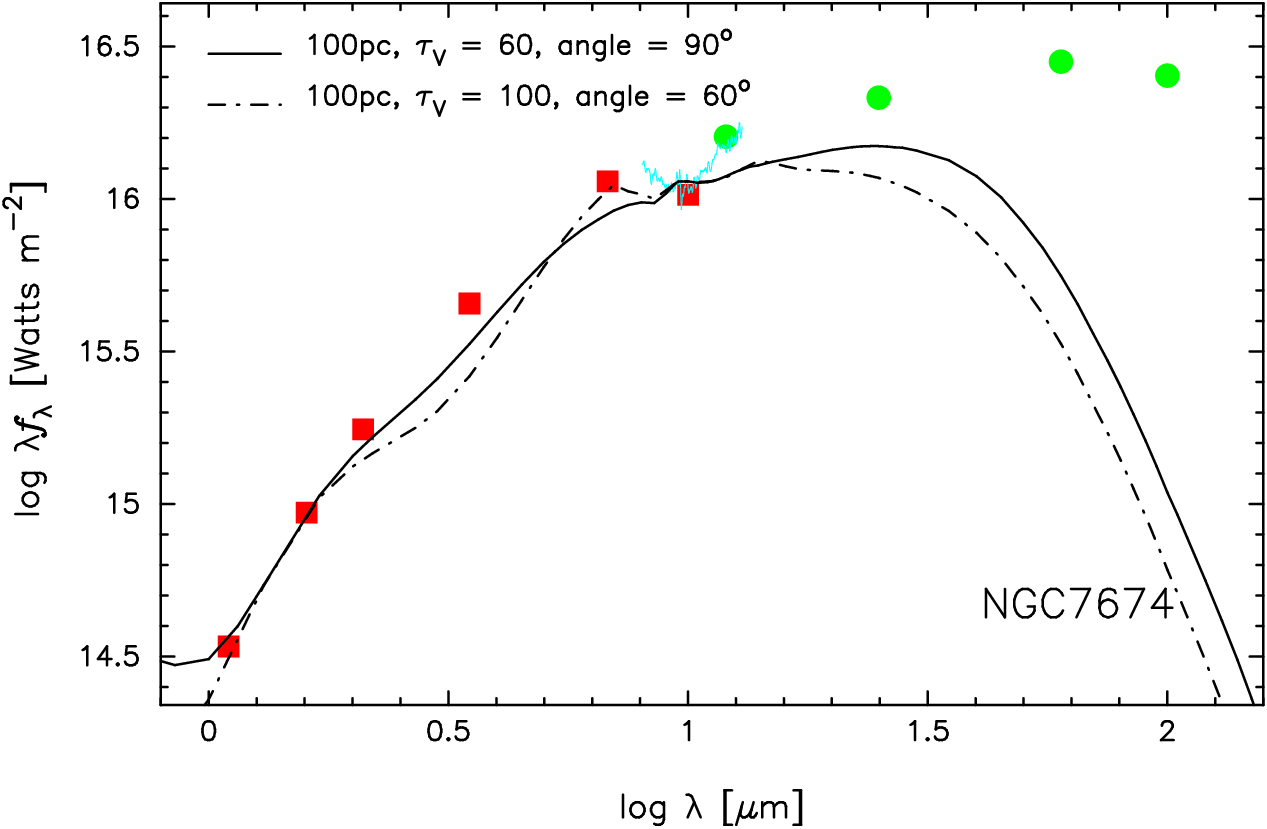}
\caption{Nuclear SEDs (squares) and N-band Michelle spectroscopy
compared to torus models (Nenkova, Ivezi\' c \& Elitzur, 2002). IRAS
data (circles), including emission from the whole galaxies, are also
shown. Some of the model parameters are shown in the figures.}
\end{figure}

These preliminary results will be greatly improved once the larger
data-base has been fully analysed. The observations and modelling will
be used to: (a) Test whether the same overall torus geometry and
physical conditions can explain the observed properties of all
objects; (b) Constrain model parameters such as optical depth,
geometry and extent of the torus; compare dust and X-ray columns for
galaxies in common with Bassani et~al.\,(1999); (c) Look for possible
correlations between torus characteristics and other traits of the
central source, such as luminosity, intermediate (ie, type S1.8 and
S1.9) classification, the presence of polarised BLR or broad wings
seen in the IR; (d) Study the circumnuclear emission to test a
`non-strict' Unified Model (Alonso-Herrero et~al., 2003; Matt, 2000)
which suggests that the obscuration in some S2s might be due to
extended material, as well as the possible connection between the lack
of hidden BLRs in cool S2s and the presence of enhanced starburst
emission (Alexander, 2001).

\begin{acknowledgments}
Support for this work was provided by Fundaci\'on Andes.
\end{acknowledgments}

\end{document}